\documentclass[aps,prl,twocolumn,superscriptaddress]{revtex4-1}

\usepackage{graphicx}

\begin{document}

%\preprint{}

\title{Incommensurate spin fluctuations in hole-overdoped superconductor KFe$_2$As$_2$}


\author{C. H. Lee}
\author{K. Kihou}
\affiliation{National Institute of Advanced Industrial Science and Technology (AIST), Tsukuba, Ibaraki 305-8568, Japan}
\affiliation{Transformative Research-Project on Iron Pnictides (TRIP), JST, Chiyoda, Tokyo 102-0075, Japan}

\author{H. Kawano-Furukawa}
\affiliation{Division of Natural/Applied Science, GSHS, Ochanomizu University, Bunkyo-ku, Tokyo 112-8610, Japan}

\author{T. Saito}
\affiliation{Department of Physics, Chiba University, Chiba 263-8522, Japan}

\author{A. Iyo}
\author{H. Eisaki}
\affiliation{National Institute of Advanced Industrial Science and Technology (AIST), Tsukuba, Ibaraki 305-8568, Japan}
\affiliation{Transformative Research-Project on Iron Pnictides (TRIP), JST, Chiyoda, Tokyo 102-0075, Japan}

\author{H. Fukazawa}
\author{Y. Kohori}
\affiliation{Transformative Research-Project on Iron Pnictides (TRIP), JST, Chiyoda, Tokyo 102-0075, Japan}
\affiliation{Department of Physics, Chiba University, Chiba 263-8522, Japan}

\author{K. Suzuki}
\affiliation{Transformative Research-Project on Iron Pnictides (TRIP), JST, Chiyoda, Tokyo 102-0075, Japan}
\affiliation{Department of Applied Physics and Chemistry, The University of Electro-Communications, Chofu, Tokyo 182-8585, Japan}

\author{H. Usui}
\affiliation{Department of Applied Physics and Chemistry, The University of Electro-Communications, Chofu, Tokyo 182-8585, Japan}

\author{K. Kuroki}
\affiliation{Transformative Research-Project on Iron Pnictides (TRIP), JST, Chiyoda, Tokyo 102-0075, Japan}
\affiliation{Department of Applied Physics and Chemistry, The University of Electro-Communications, Chofu, Tokyo 182-8585, Japan}

\author{K. Yamada}
\affiliation{WPI, Tohoku University, Sendai 980-8577, Japan}

%\thanks{}


%\date{\today}

\begin{abstract}
A neutron scattering study of heavily hole-overdoped superconducting KFe$_2$As$_2$ revealed a well-defined low-energy incommensurate spin fluctuation at [$\pi(1\pm2\delta$),0] with $\delta$ = 0.16.  
The incommensurate structure differs from the previously observed commensurate peaks in electron-doped $A$Fe$_2$As$_2$ ($A$ = Ba, Ca, or Sr) at low energies.  
The direction of the peak splitting is perpendicular to that observed in Fe(Te,Se) or in Ba(Fe,Co)$_2$As$_2$ at high energies.  
A band structure calculation suggests interband scattering between bands around the $\Gamma$ and X points as an origin of this incommensurate peak.  
The perpendicular direction of the peak splitting can be understood 
within the framework of multiorbital band structure.  
The results suggest that spin fluctuation is more robust in hole-doped than in electron-doped samples, which can be responsible for the appearance of superconductivity in the heavily hole-doped samples.  
\end{abstract}


%\keywords{}

\maketitle

It is widely believed that studying magnetic interaction is quite essential to understanding high-transition temperature (high-$T_c$) superconductivity in both copper oxides and iron pnictides, in which superconductivity emerges as a consequence of the suppression of long-range antiferromagnetic (AFM) order.  For this purpose, neutron scattering techniques have played significant roles since they can determine both the wave vector ($\bf q$) and the energy dependence of the magnetic excitations.  

%=========================================================
\begin{figure}
\includegraphics[width=\columnwidth]{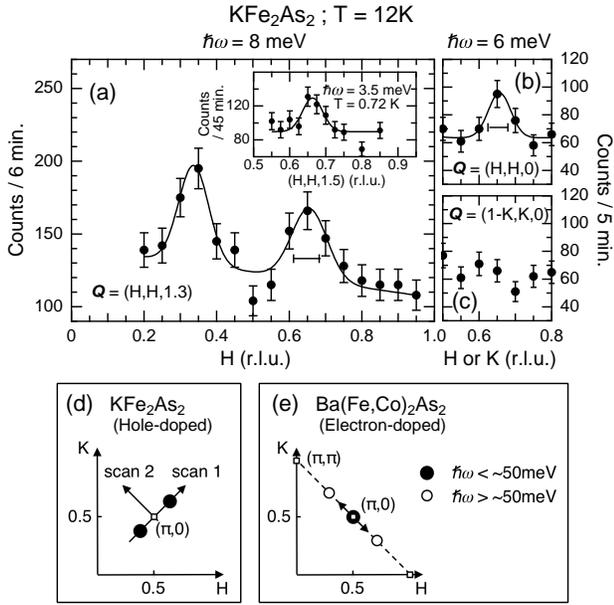}
\caption{\label{incomme} (a)-(c) Constant-energy scan of KFe$_2$As$_2$ at $T$ = 12 K along (a) (H,H,1.3) at $\hbar\omega$ = 8 meV, (b) (H,H,0) at $\hbar\omega$ = 6 meV, and (c) (1-K,K,0) at $\hbar\omega$ = 6 meV.  Horizontal bars depict the instrumental $q$ resolution.  Scan trajectories are indicated by arrows in (d) as scan 1 for (a,b) and scan 2 for (c).  
The spectrum in (a) is obtained in the (H,H,L) zone, and those in (b,c) are obtained in the (H,K,0) zone.  Solid lines indicate Gaussian fits.  
The inset in (a) is the $\bf q$ spectrum at $\hbar\omega$ = 3.5 meV and $T$ = 0.72 K observed by using HER spectrometer.  
(d,e) Schematic illustrations of magnetic peak positions of (d) KFe$_2$As$_2$ and (e) Ba(Fe,Co)$_2$As$_2$.}
\end{figure}
%=========================================================

As for the Fe-based superconductors, the outcome of the neutron scattering studies 
\cite{Cruz2008,Huang2008,Lumsden2009,Lester2010,Li2010,Chi2009,Lumsden2010,SHLee,Christianson2008} 
is summarized as follows.  
Nondoped $R$FeAsO ($R$ = rare earth) and $A$Fe$_2$As$_2$ ($A$ = Ba, Ca, or Sr) exhibit long-range AFM order, 
where a commensurate magnetic peak is observed at $\bf q$ = ($\pi$,0) \cite{Cruz2008,Huang2008}.  
For the superconducting samples of electron-doped $A$Fe$_2$As$_2$, 
short-range and commensurate spin fluctuations are observed at the same $\bf q$ position 
of ($\pi$,0) at low energies [Fig. \ref{incomme}(e)] \cite{Lumsden2009,Lester2010,Li2010,Chi2009}.  
With increasing energy, the magnetic peaks of Ba(Fe,Co)$_2$As$_2$ split into two peaks 
towards ($\pi$,$\pi$) above $\hbar\omega$ $\sim$ 50 meV because of anisotropic dispersion \cite{Lester2010,Li2010}.  
In superconducting Fe(Se,Te), on the other hand, incommensurate spin fluctuations are observed around ($\pi$,0), 
where the direction of the peak splitting is equivalent to that of Ba(Fe,Co)$_2$As$_2$ \cite{Lumsden2010,SHLee}.  

Broadly speaking, there are two classes of proposals to account for the experimental observation.  
Based on the localized model, AFM order arises from the local moments of Fe sites described with a Heisenberg Hamiltonian.  
On the itinerant model, AFM order is a consequence of the nesting of Fermi surfaces 
\cite{Graser,Yaresko,Knolle,Kuroki2008,Ikeda,Park2010}.  
At this moment, the microscopic origin of the AFM order is still controversial.  

In general, the complexity can be reduced in the heavily doped region.  
KFe$_2$As$_2$, which is the end member of the Ba$_{1-x}$K$_x$Fe$_2$As$_2$ system, 
seems to be a canonical Fermi liquid as evidenced by the T$^2$ dependence of the resistivity 
and by the fact that it follows the Kadowaki-Woods relationship \cite{Hashimoto}.  
Moreover, angle-resolved photoemission spectroscopy (ARPES) and de Haas-van Alphen (dHvA) measurements have shown 
that its band structure is mostly consistent with that predicted by the band structure calculation \cite{Sato,Terashima2010}.  
These suggest that the itinerant model can describe the physics of KFe$_2$As$_2$, 
which offers the starting point in exploring the phase diagram of the Fe-based superconductors from the overdoped side.

KFe$_2$As$_2$ is also an intriguing material in itself.  
The superconducting phase of Ba$_{1-x}$K$_x$Fe$_2$As$_2$ remains even in the 50 \% hole-doped KFe$_2$As$_2$ ($T_c$ = 3.4 K).  
This is in sharp contrast to electron-overdoped samples of Ba(Fe$_{1-x}$Co$_x$)$_2$As$_2$ (x = 0.24) and 
LaFeAsO$_{1-x}$F$_x$ (x = 0.158), where spin fluctuations as well as superconductivity disappear quickly \cite{Wakimoto,Matan2010}.  
The superconductivity is suggested to be unconventional from the indication of a nodal superconducting gap 
\cite{Fukazawa2009,Dong2010,Hashimoto}.  
Furthermore, the large Sommerfeld constant ($\gamma$ = 93 mJ/molK$^2$) \cite{Fukazawa2011} 
suggests strong electron correlation, and 
nuclear magnetic resonance measurements indicate the existence of spin fluctuation \cite{Zhang,Fukazawa2009} 
although there is no interband nesting in KFe$_2$As$_2$, 
pointing towards the superconductivity associated with magnetism.  

Thus, we conducted inelastic neutron scattering measurements 
and a band structure calculation of KFe$_2$As$_2$.  
We found that spin fluctuations still occur in heavily hole-overdoped samples and are incommensurate, 
in contrast to the commensurate magnetic peaks observed in electron-doped $A$Fe$_2$As$_2$.  
We also found that the band structure calculation provides a quantitatively correct description of the experimental results.  

High-quality single crystals of KFe$_2$As$_2$ were grown by the self-flux method, which is described in detail elsewhere \cite{Kihou}.  
The $T_c$ of the grown single crystals was determined to be 3.4 K with a transition width of 0.2 K from the temperature dependence of zero-field-cooled magnetization \cite{Kihou}.  
The $T_c$ of individual crystals varied by less than 0.2 K.  
For inelastic neutron scattering measurements, approximately 300 tabular-shaped single crystals ($\sim$0.4 cm$^3$)  were coaligned on a thin Al sample holder. 
The total mosaic spreads of the coaligned samples were $\sim$4.5$^\circ$ and $\sim$2$^\circ$ (FWHM) in the (H,H,L) and (H,K,0) scattering planes, respectively.  

%=========================================================
\begin{figure}
\includegraphics[width=\columnwidth]{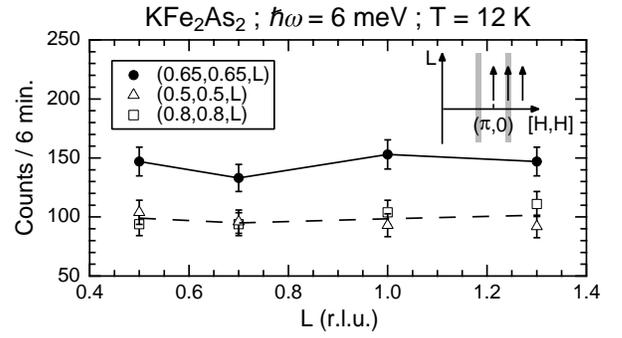}
\caption{\label{rod} $L$ dependence of the magnetic peak [$\bf Q$ = (0.65,0.65,$L$)] and background [$\bf Q$ = (0.5,0.5,$L$) and (0.8,0.8,$L$)] intensities at $T$ = 12 K for $\hbar\omega$ = 6 meV.  Inset shows the scan trajectories.}
\end{figure}
%=========================================================

%=========================================================
\begin{figure}
\includegraphics[width=\columnwidth]{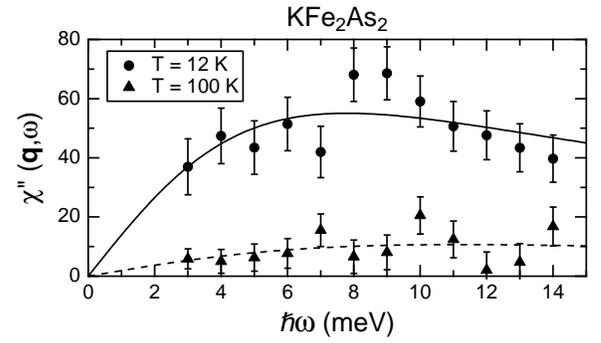}
\caption{\label{kai} Energy spectrum of $\chi^{\prime \prime}({\bf q},\omega)$ at $T$ = 12 and 100K.  
The solid and dashed lines are fits obtained using Eq. (1) at $T$ = 12 and 100 K, respectively.}
\end{figure}
%=========================================================

Inelastic neutron scattering measurements were conducted at JRR-3 of the Japan Atomic Energy Agency by using the triple-axis spectrometers TOPAN, GPTAS, and HER.  
The final neutron energy was fixed at $E_{f}$ = 14.8 meV in TOPAN and GPTAS by using 
vertical focusing pyrolytic graphite (PG) crystals as a monochromator and analyzer.  
In HER, the final neutron energy was fixed at $E_{f}$ = 5.3 meV by using vertical and double focusing PG crystals as a monochromator and analyzer, respectively.  
The horizontal collimator sequences were open-100'-S-60'-open, open-80'-S-80'-80', and guide-open-S-radial-open 
in TOPAN, GPTAS, and HER, respectively, where S denotes the sample position.  
A sapphire crystal in TOPAN and a PG filter in TOPAN, GPTAS, and HER were inserted to remove higher-order and high-energy neutrons.  
Samples were aligned in the (H,K,0) zone in GPTAS and in the (H,H,L) zone in TOPAN and HER.  
A closed cycle $^4$He and $^3$He refrigerator was used to cool samples down to 12 and 0.72 K, respectively.  

Well-defined incommensurate spin fluctuations of KFe$_2$As$_2$ were found at $\bf Q$ = (0.5$\pm\delta$,0.5$\pm\delta$,$L$), 
which corresponds to the [$\pi(1\pm2\delta$),0] position in the ab plane.  
Figure \ref{incomme}(a) shows the incommensurate $\bf q$ spectrum with incommensurability $\delta$ = 0.16 at $\hbar\omega$ = 8 meV and $T$ = 12 K in the (H,H,L) zone.  
The incommensurate peaks are also confirmed at $\hbar\omega$ = 3.5 meV [inset of Fig. \ref{incomme}(a)], 6, 10, and 12 meV.  
To determine whether there is also a peak in the perpendicular direction, we measured in the (H,K,0) zone 
with the scan trajectories labeled scan 1 and 2 in Fig. \ref{incomme}(d).  
A well-defined peak was again observed at $\bf Q$ = (0.66,0.66,0) in scan 1 [Fig. \ref{incomme}(b)] 
but not in the perpendicular scan [scan 2, Fig. \ref{incomme}(c)].  
Note that the present well-defined incommensurate peaks differ from the commensurate peaks 
observed in electron-doped $A$Fe$_2$As$_2$ at low energies.  
Magnetic peaks of Ba(Fe,Co)$_2$As$_2$ only split above $\hbar\omega$ $\sim$ 50 meV because of anisotropic dispersion \cite{Lester2010,Li2010}.  
In addition, the direction of the present peak splitting is perpendicular to that observed in 
Ba(Fe,Co)$_2$As$_2$ and Fe(Se,Te) [Fig. \ref{incomme}(e)] \cite{Lester2010,Li2010,Lumsden2010,SHLee}.  
Figure \ref{rod} shows the $L$ dependence of the intensity at $\hbar\omega$ = 6 meV and $T$ = 12 K with different $H$ positions.  
The net intensity is almost constant in the $\bf Q$ range of 0.5 $\leq$ $L$ $\leq$ 1.3 r.l.u., 
indicating that the incommensurate spin fluctuation is two dimensional.  

The energy dependence of the net intensity of the incommensurate peaks at $T$ = 12 and 100 K was determined by measuring the intensity 
at $\bf Q$ = (0.66,0.66,1.5) and (0.8,0.8,1.5) as the peak and background intensities, respectively.  
The dynamical magnetic susceptibility $\chi^{\prime \prime}({\bf q},\omega)$ was obtained by multiplying the net intensity by 
$[1-\exp(-\hbar\omega$/k$_{\rm B}T)]$, where k$_{\rm B}$ denotes the Boltzmann constant.  
Figure \ref{kai} shows the energy dependence of $\chi^{\prime \prime}({\bf q},\omega)$ at $T$ = 12 and 100 K 
in the energy range of 3 $\leq$ $\hbar\omega$ $\leq$ 14 meV.  
At $T$ = 12 K, $\chi^{\prime \prime}({\bf q},\omega)$ increases with increasing energy up to $\hbar\omega$ $\sim$ 9 meV 
and then slowly decreases.  
Upon heating from $T$ = 12 to 100 K, $\chi^{\prime \prime}({\bf q},\omega)$ is suppressed over the entire energy range, 
ensuring that contamination from phonons is quite small in the observed magnetic signals.  
We fitted the energy dependence of $\chi^{\prime \prime}({\bf q},\omega)$ using a phenomenological function applicable to 
correlated spin systems in Fermi liquids without magnetic long-range ordering,

\begin{equation}
\chi^{\prime \prime}({\bf q},\omega) = \chi_0\frac{\Gamma\hbar\omega}{\Gamma^2+\hbar\omega^2} \ \ ,
\end{equation}

\noindent where $\chi_0$ represents the strength of the antiferromagnetic correlation and $\Gamma$ is the damping constant.  
Fitting results are depicted by solid and dashed lines in Fig. \ref{kai}, which represent the data reasonably well.  
The evaluated values are $\Gamma$ = 7.8$\pm$1 meV, $\chi_0$ = 220$\pm$11 at $T$ =12 K, 
and $\Gamma$ = 11$\pm$6 meV, $\chi_0$ = 43$\pm$9 at $T$ = 100 K.  

The present results demonstrate that spin fluctuations clearly exist even in heavily overdoped KFe$_2$As$_2$.  
Usually spin fluctuations that are remnants of antiferromagnetic ordering are expected to be damped 
in overdoped samples because of carrier doping.  
The disappearance of the Fermi surface nesting between hole and electron pockets in KFe$_2$As$_2$ strengthens this expectation.  
In fact, spin fluctuations disappear in electron-overdoped Ba(Fe$_{1-x}$Co$_x$)$_2$As$_2$ (x = 0.24) 
and LaFeAsO$_{1-x}$F$_x$ (x=0.158) samples \cite{Wakimoto,Matan2010}.  
Contrary to the expectation, however, spin fluctuations have been found in 50 \% hole-doped KFe$_2$As$_2$, 
although the doped carrier concentration is more than twice that in electron-overdoped samples whose spin fluctuations disappear.  
This suggests that spin fluctuation is more robust in hole-doped samples than in electron-doped samples, 
which can be responsible for the appearance of superconductivity in KFe$_2$As$_2$.  

To investigate whether the present incommensurate spin fluctuations could originate from Fermi surface nesting, 
we compared the wave vector $\bf q$ of the present magnetic peaks with the possible nesting vector in KFe$_2$As$_2$.  
According to ARPES and dHvA measurements, hole pockets exist around the $\Gamma$ point, 
where intraband nesting can occur in the [110] direction with a nesting vector of (0.2,0.2,0) or (0.24,0.24,0) \cite{Sato,Terashima2010}.  
However, the observed magnetic peaks locate at $\bf q$  = (0.34,0.34,0), which is far from the nesting vector within the hole pocket.  

The other possibility is interband scattering between a hole pocket around the $\Gamma$ point and a flat band around the X point 
located just above the Fermi energy.  
Although no electron Fermi surface exists around the X point, interband scattering can occur with finite energy transfer
resulting from inelastic neutron scattering.  
The corresponding scattering vector estimated from the band structure could be (0.38,0.38,0) or (0.4,0.4,0) \cite{Sato,Terashima2010}, 
which is reasonably close to the wave vector of the present magnetic peaks.  

\begin{figure}
\includegraphics[width=\columnwidth]{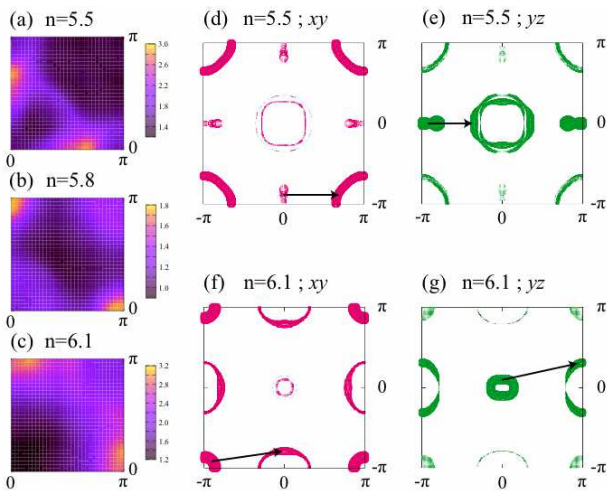}
\caption{\label{theory} Contour plots of the spin susceptibility for 
(a) n=5.5, (b) $n=5.8$, and 
(c) $n=6.1$.  The 
strength of the $xy$ and $yz$ orbital character on the Fermi surface for 
(d),(e) $n=5.5$ and (f),(g) $n=6.1$. The arrows connect Fermi surface 
portions with similar orbital character.  
No arrow toward the inner hole Fermi surface is depicted in (e) since this Fermi surface
has a strong three-dimensional nature in the orbital character and its contribution to the spin fluctuation is small.  
}
\end{figure}

The validity of the interband scattering picture was 
examined using a theoretical calculation 
based on a random phase approximation applied to the 
five-orbital model of KFe$_2$As$_2$ obtained from a first 
principles calculation using maximally localized Wannier orbitals.  
In $A$Fe$_2$As$_2$, the Brillouin zone (BZ) unfolding 
process that adopts the  
unit cell with one Fe per unit cell \cite{Kuroki2008} 
cannot be strictly done, but here we have adopted a model obtained by approximately 
unfolding the BZ. Although this model does not reproduce the original band
structure perfectly, the portions of the Fermi surface 
relevant to the spin fluctuations are well reproduced. 
We have also confirmed that 
the spin susceptibility obtained from the five-orbital model agrees with that 
obtained from the ten-orbital model (with two Fe per unit cell) \cite{Suzuki2-2010} 
when both of them are presented in the same folded BZ.  
Here, we took $64\times 64\times 16$ $k$-point meshes, 512 Matsubara
frequencies and a temperature of $T=0.04$ eV.
The calculated spin susceptibility is shown in Fig. \ref{theory} (a) with the unfolded BZ.  
The peak is at [$\pi(1\pm2\delta$),0] with $\delta$ = 0.17, 
corresponding to $\bf q$ = (0.33,0.33,0), which  
is in striking agreement with the experiment.
The band filling of KFe$_2$As$_2$ is $n=5.5$ (5.5 electrons per Fe), but 
we also hypothetically changed the band filling to 
$n=5.8$ (corresponding to Ba$_{0.6}$K$_{0.4}$Fe$_2$As$_2$). 
As shown in Fig. \ref{theory} (b), the peak moves to the commensurate 
position $(\pi,0)$ for $n=5.8$.  In between the band filling 5.8 and 5.5, the 
peak moves continuously from the commensurate to the incommensurate 
position toward (0,0) with decreasing $T_c$.  
This continuous variation indicates that the magnetic 
peak has the same origin in the entire (Ba,K)Fe$_2$As$_2$ system,
namely, interband scattering.  

%=========================================================
%=========================================================

We also performed a similar calculation for $n=6.1$, the 
electron-doped case, using a model for BaFe$_2$As$_2$.  
As shown in Fig. \ref{theory}(c), the peak moves toward $(\pi,\pi)$ in the unfolded BZ.  
Considering that the present calculated spin susceptibility is an integral of $\chi^{\prime \prime}({\bf q},\omega)$ in a wide energy range, 
it is consistent with the experimental results of Ba(Fe,Co)$_2$As$_2$ at high energies.  
The difference between the electron- and hole-doped cases cannot be explained 
by the distribution of the density of states without taking into account the orbital character.  
Figures \ref{theory}(d)-\ref{theory}(g) show the strength of the 
orbital character on the Fermi surfaces of 
KFe$_2$As$_2$ $(n=5.5)$ and BaFe$_2$As$_2$ with $n=6.1$ in the unfolded BZ.  
Here we plot 
the states within a finite energy range $-\Delta E< E(k)-E_F <\Delta E$ 
with $\Delta E=0.02$eV, which can contribute to the spin fluctuations.  
Although Fermi surfaces near the wave vector 
$(\pi,0)/(0,\pi)$ are barely present in KFe$_2$As$_2$, there are the states 
in the vicinity of the Fermi level 
originated from a nearly flat band lying
close to the Fermi level.
The spin fluctuations develop at wave vectors that bridge the portions of 
the Fermi surface having similar orbital character \cite{KKheight}. 
Interestingly, the wave vectors connecting the $xy$ orbital portions and those connecting 
the $yz$ (or $xz$, not shown) 
portions nearly coincide, resulting in spin fluctuations at the same $\bf q$ position in each electron- and hole-doped cases.  
The wave vector deviates from $(\pi,0)$ toward $(\pi,\pi)$ in the electron-doped case, 
while it deviates toward $(0,0)$ in the hole-doped case.
In this view, the difference of the magnetic peak structure 
between the electron- and hole-doped cases can originate from the 
multiorbital nature of the system.

%Perpendicular direction of the magnetic peak splitting in Ba(Fe,Co)$_2$As$_2$ at high energies comparing with 
%the present peaks can be explained by both the itinerant and localized model.  
%Theoretical calculations based on the itinerant model indicate that magnetic peaks can appear at ($\pi$,$\delta\pi$) 
%in electron-doped samples \cite{Graser,Yaresko,Knolle,Kuroki2008,Ikeda} consistent with experimental results.  
%It has been proposed that sign of a carrier can be responsible for the direction of the peak splitting 
%because of different $\bf q$ distribution of the density of states near the Fermi level between electron- and hole-doped samples.  
%Based on the localized model, the splitting can be 
%reproduced using effective nearest-neighbour and next-nearest-neighbour exchanged coupling constants \cite{Lester2010}.  
%The successful reproduction in Ba(Fe,Co)$_2$As$_2$ reminds the possibility 
%that the local model is also applicable to the spin fluctuation of KFe$_2$As$_2$, 
%although the itinerant model can explain its incommensurate peak successfully.  

In summary, well-defined incommensurate spin fluctuations have been observed at the [$\pi(1\pm2\delta$),0] position 
with $\delta$ = 0.16 in heavily hole-overdoped KFe$_2$As$_2$ by an inelastic neutron scattering technique.  
The results differ from the commensurate peaks observed in electron-doped $A$Fe$_2$As$_2$ at low energies.  
In addition, the direction of the present peak splitting is perpendicular to that previously observed in other Fe-based superconductors.  
The agreement with theoretically calculated peak positions suggests that the spin fluctuations originate from interband scattering 
between a hole pocket around the $\Gamma$ point and a band around the X point.  
The perpendicular direction of the peak splitting between electron- and hole-doped $A$Fe$_2$As$_2$ can be understood 
within the framework of multiorbital band structure.  
Robust spin fluctuations can be responsible for the appearance of superconductivity in heavily hole-doped samples.  

\begin{acknowledgments}
We thank E. M. Forgan for scientific discussions and Y. Nakano and H. Kikuchi for their technical help.  
This study was supported by a Grant-in-Aid for Scientific Research on Innovative Areas 
``Heavy Electrons" (No. 20102005 and No. 21102505) from the Ministry of Education, Culture, Sports, Science and Technology of Japan 
and by Grants-in-Aid for Scientific Research A (No. 22244039) and C (No. 22540380) 
from the Japan Society for the Promotion of Science.  
\end{acknowledgments}

\end{document}